# Quantum Transport in Air-stable Na$_3$Bi Thin Films

*Chang Liu[†,‡,§], Golrokh Akhgar[†,§,∥], James L. Collins[†,‡,§], Jack Hellerstedt[†,‡], Shaffique Adam[⊥,#,¶],*

*Michael S. Fuhrer[\*,†,‡,§] and Mark T. Edmonds[\*,†,‡,§]*

[†]School of Physics and Astronomy, Monash University, Victoria 3800 Australia

[‡]Monash Centre for Atomically Thin Materials, Monash University, Victoria 3800 Australia

[§]ARC Centre of Excellence in Future Low-Energy Electronics Technologies, Monash University, Victoria 3800 Australia

[∥]School of Science, RMIT University, Melbourne, Victoria 3001 Australia

[⊥]Yale-NUS College, 16 College Avenue West, 138527 Singapore

[#]Centre for Advanced 2D Materials, National University of Singapore, 6 Science Drive 2, 117546 Singapore

[¶]Department of Physics, National University of Singapore, 2 Science Drive 3, 117551 Singapore

# ABSTRACT

Na$_3$Bi has attracted significant interest in both bulk form as a three-dimensional topological Dirac semimetal and in ultra-thin form as a wide-bandgap two-dimensional topological insulator. Its extreme air sensitivity has limited experimental efforts on thin- and ultra-thin films grown via molecular beam epitaxy to ultra-high vacuum environments. Here we demonstrate air-stable Na$_3$Bi thin films passivated with magnesium difluoride (MgF$_2$) or silicon (Si) capping layers. Electrical measurements show that deposition of MgF$_2$ or Si has minimal impact on the transport properties of Na$_3$Bi whilst in ultra-high vacuum. Importantly, the MgF$_2$-passivated Na$_3$Bi films are air-stable and remain metallic for over 100 hours after exposure to air, as compared to near instantaneous degradation when they are unpassivated. Air stability enables transfer of films to a conventional high-magnetic field cryostat, enabling quantum transport measurements which verify that the Dirac semimetal character of Na$_3$Bi films is retained after air exposure.

INTRODUCTION

Na$_3$Bi is a three-dimensional (3D) topological Dirac semimetal (TDS) in bulk form, with linear energy dispersion in 3D momentum space.[1-2] These 3D Dirac fermions provide opportunities to study the chiral anomaly,[3] Dirac-point physics due to ultra-low charge puddling,[4] and ambipolar transport behaviour using electric-field effect devices.[5] Perhaps the most promising attribute of Na$_3$Bi is that when confined to atomic thinness, it becomes a large bandgap two-dimensional (2D) topological insulator, that undergoes an electric field induced quantum phase transition from topological to trivial insulator.[6] This is particularly promising in the development of low-energy switches and the creation of a topological transistor.[6-8]

Yet to date, the extreme air sensitivity of Na$_3$Bi devices (lasting only seconds in atmosphere) has limited the study of Na$_3$Bi thin and ultra-thin films to ultra-high vacuum (UHV) environments. In order to fully exploit the electronic properties of Na$_3$Bi, like the fabrication of air-stable devices, the development of inert passivation layers for ambient conditions is crucial.

In this study, we utilize magnesium difluoride (MgF$_2$) and silicon (Si) as passivation layers in order to prevent the in-air degradation of epitaxial (MBE) Na$_3$Bi thin films. We have previously demonstrated the chemical neutrality of Si when used as a substrate to grown Na$_3$Bi.[6] The MgF$_2$ was chosen for its demonstrated effectiveness as an optically transparent capping layer for the topological insulator Bi$_2$Se$_3$.[9-10] We demonstrate that 20 nm Na$_3$Bi films remain stable in air for around 2 hours with Si capping and over 100 hours with MgF$_2$ capping, allowing electrical characterization of the films at large magnetic fields across a wide temperature range. Electrical measurements reveal that whilst the passivation layers keep Na$_3$Bi metallic in air, a significant amount of disorder is induced that lowers the overall mobility. This is possibly due to non-uniform and imperfect encapsulation at the edges of the sample. However, quantum transport measurements show nearly perfect weak anti-localization, verifying that spin-momentum locking, the hallmark of Dirac semimetals, is retained by Na$_3$Bi films after air-exposure. These results thus represent an important first step in achieving air-stable Na$_3$Bi Dirac semimetal devices.

EXPERIMENTAL METHODS

Na$_3$Bi 20 nm films are grown via MBE on Al$_2$O$_3$[0001] substrates using a two-step growth method[11] where the first 2 nm of Na$_3$Bi was grown at 120 °C for nucleation, the substrate temperature was subsequently ramped up to 330 °C over the next 5 nm of growth, and the remaining 13 nm are grown at 330 °C. An extra 10 min of Na-flux only annealing was applied at 330 °C before cooling down to optimize the crystal quality. For these experiments, the UHV device fabrication and capping scheme is shown in Figure 1a-d. Specifically, we used a shadow mask with Hall bar geometry attached to the substrate during the Na$_3$Bi film growth. The shadow mask is removed in UHV after growth which allows the deposited Si or MgF$_2$ capping layer to completely cover the entire Hall bar. For the transport measurements, for each temperature $T$ and value of magnetic field $B$, a DC excitation is applied to pass through the main channel of the Hall bar, and the longitudinal and transverse voltages between the side arms of the Hall bar are recorded as $V_{xx}$ and $V_{xy}$ respectively.

RESULTS & DISCUSSION

We first study the transport properties of 20 nm $Na_3Bi$ thin films in UHV of as-grown and capped films. The samples were transferred in UHV to the attached analysis chamber which is equipped with a ±1 T superconducting magnet at 5 K. Figure 1e shows the Hall carrier density and mobility as grown and after capping at 5 K in UHV. The as grown 20 nm $Na_3Bi$ films are *n*-type with carrier densities and mobilities in the range of $1.9 \times 10^{17}$ – $4.3 \times 10^{18}$ cm$^{-3}$ and 2000 - 6000 cm$^2$/Vs respectively, similar to previously reported values.[11] The two films capped with Si show a slightly reduced *n*-type density, whilst both small increases and decreases in density are observed with $MgF_2$ capping. Both Si and $MgF_2$ capping cause moderate reductions in mobility, with the mobility consistently ~2000 cm$^2$/Vs for all samples after capping. These changes to doping and mobility upon capping suggest that $MgF_2$ or Si do not chemically react with $Na_3Bi$. This is consistent with the previous observation of inertness of the $Na_3Bi$-Si interface using X-ray photoelectron spectroscopy (XPS).[6]

Figure 2 shows the in-air stability test conducted by recording the sheet resistance, $R_{sheet}$, from when samples were removed from the UHV chamber into ambient conditions. The process of transferring sample from the UHV chamber into air takes several minutes, which includes $N_2$ chamber venting and the in-air resistivity measurement set up. The sheet resistance of uncapped $Na_3Bi$ (red line in Figure 2) increases from $10^2$ Ω to above GΩ during the time it takes to commence in-air resistance measurements, suggesting catastrophic degradation of $Na_3Bi$ upon air exposure. In Figure 2, we also show the resistance in UHV and after air exposure of two 20 nm $Na_3Bi$ films that have been passivated with Si (green line) and $MgF_2$ (blue line). Both passivated films show an order of magnitude increase in $R_{sheet}$ (from ~$10^2$ Ω to ~$10^3$ Ω) during the transfer from UHV to atmosphere. The increase in $R_{sheet}$ could result from inhomogeneity of the capping layer with micro-cracks or pin holes, or an issue at the edge of the Hall bar, as Si and MgF$_2$ are amorphous and do not uniformly deposit on the Na$_3$Bi surface. The slower rise in sheet resistance in Figure 2 demonstrates that the MgF$_2$ capping layer has a better performance compared to the Si capping layer. The Na$_3$Bi thin film shows a stable and slow-degradation when it is capped with the MgF$_2$, remaining metallic for over 100 hours. The Si-capped film exhibits faster initial degradation rate and shows a significant increase after just 2 hours of air exposure, becoming completely insulating after 10 hours.

Figure 3a shows the magnetoresistance curves after capping the Na$_3$Bi thin films using MgF$_2$ and transferring in-air from UHV chamber to a low temperature cryostat (Quantum Design PPMS). The film shows, in Figure 3a, a strong positive transverse magnetoresistance, $\frac{\Delta R}{R}(B) = [R(B) - R(B=0)]/R(B=0)$. In order to understand the origin of the magnetoresistance, we fit our low-field quadratic magnetoresistance data to the semiclassical Drude formalism:[11]

$$\rho_{xx}(B) = \rho_{xx}(B=0)[1 + A(\mu B)^2] \qquad (1)$$

where $A$ is a dimensionless coefficient.

Figure 3b shows the temperature dependence of the coefficient $A$ from fits of the data in Figure 3a to Eqn. 1. At high temperatures, $A$ saturates to a temperature-independent value, suggesting that magnetoresistance in this temperature regime is a classical effect. Quadratic low-field magnetoresistance has been shown to be a generic aspect of classical transport in a disordered conductor.[5,12] However, theory predicts that $0 < A \leq 1/2$, and experimentally $A \lesssim 1$,[5,11-12] while we observe that $A$ saturates to a value around 850 above 50 K, which is superficially at odds with classical magnetoresistance in a disordered system. The mobility aslo appears unphysically low; it corresponds to a mean free path of 0.8 Å, smaller than the lattice constant, violating the Ioffe-Regel limit.

However, disorder has unusual implications in Dirac electronic systems. At low carrier density the disorder potential can create both electron-like and hole-like local regions or puddles, with no energy gap of transport between the electron and hole puddles.[13-14] Charge pudding has been observed in graphene,[14] where it is strongly dependent on the underlaying substrate, with large fluctuations in Fermi energy $\Delta E_F$ ~ 100 meV on $SiO_2$ [15] and much lower $\Delta E_F = 3 - 5$ meV on $h$-BN [16]. Pudding has also been observed in the 2D surface state of 3D topological insulators,[17] and in 3D topological Dirac semimetals, such as $Na_3Bi$ thin films on sapphire in UHV [4] where $\Delta E_F$ ~ 3 – 5 meV comparable to graphene on $h$-BN. However, $\Delta E_F$ for $Na_3Bi$ films is increased when grown on more disordered substrates such as $SiO_2$.[5]

Similarly, electron-hole puddling has specific consequences for electronic transport in a Dirac system. In particular, the Hall coefficient $R_H$ as a function of average carrier density $n$ is no longer single valued; it tends to zero as $R_H$ ~ $1/n$ at very large carrier density, and at very small carrier density $R_H$ ~ $n$ where the electron and hole contributions nearly cancel, with a maximum at $n \approx n^*$, the characteristic carrier density in the puddles. The conductivity is also altered significantly, saturating to a minimum conductivity $\sigma_{min} = n^* e\mu$ for $|n| < n^*$.[13] In the puddle regime the Hall effect overestimates the carrier density, and thus the Hall mobility underestimates the true mobility. This suggests that our unphysically large $A$ in Eqn. 1 results from an underestimation of the mobility $\mu$. If we assume instead that our sample is in the puddle regime where $A \sim 0.5$ and $\sigma = \sigma_{min} = n^* e \mu_{corr}$, then we find a corrected mobility $\mu_{corr} = 80.2$ cm$^2$/Vs, and $n^* = 1.1\times10^{19}$ cm$^{-3}$. The puddle carrier density $n^*$ is about 50 times higher for air-exposed Na$_3$Bi than observed in Na$_3$Bi on SiO$_2$ in vacuum,[5] implying a 50-fold increase in impurity density, as $n^*$ is proportional to impurity density for a Dirac system. This is consistent with our observation of mobility about 40 times lower for air-exposed Na$_3$Bi compared to Na$_3$Bi on SiO$_2$. The inset to Figure 3b summarizes the effect of pudding. The measured carrier density from the Hall effect is 4-5×10$^{20}$ cm$^{-3}$ (blue filled square) corresponds to a mobility of ~2 cm$^2$/Vs after in-air transfer. Assuming that the sample is in the electron-hole puddle regime, the corrected values are $\mu_{corr} = 80.2$ cm$^2$/Vs, and $n^* = 1.1\times10^{19}$ cm$^{-3}$, indicated by the open blue square.

At temperatures below 50 K the magnetoresistance is larger and becomes cusp-like, suggesting quantum corrections to the conductivity. Figure 4a shows the magnetoconductance, $\Delta\sigma_{xx} = [\sigma_{xx}(B) - \sigma_{xx}(B = 0)]/(e^2/h)$, at various temperature 2 – 20 K, in units of $e^2/h$, where $e$ is the elementary charge and $h$ is the Planck's constant. The classical component of the magnetoresistance (Eqn. 1) has been subtracted from the data before inverting the resistivity tensor, though the correction is small for $T \leq 20$ K (WAL dominates the magnetoconductance) and makes a negligible difference in the fit parameters in the analysis below. We fit the data to the reduced Hikami-Larkin-Nagaoka formula (H.L.N.)[18] in the limit of perfect spin-orbit coupling in the low field region:

$$\frac{\Delta\sigma_{xx}}{e^2/h} = -\frac{\alpha}{2\pi}\left[\ln\left(\frac{B_\phi}{B}\right) - \Psi\left(\frac{1}{2}+\frac{B_\phi}{B}\right)\right] \quad (2)$$

where $\alpha$ is a pre-factor signifying the type of localization as 1/2 for WAL and -1 for WL, the fitting parameter $B_\phi$ is the phase coherence field which relates to the phase coherence length $L_\phi = \sqrt{\frac{h}{8\pi e B_\phi}}$. The fits are shown as black solid lines in Figure 4a, where $\alpha$ and $B_\phi$ are fit parameters. Equation 2 provides an excellent fit to the data at all temperatures, indicating that air-exposed $Na_3Bi$ still retains near-perfect spin-momentum locking as expected for a Dirac semimetal, even though disorder is increased. Values of $\alpha$ range from 0.37 – 0.5 as a function of $T$ as shown in Figure 4b, close to the theoretical value of 1/2. The discrepancy could result from microscopic (electron-hole pudding) or macroscopic (induced by air exposure) inhomogeneity in the sample.

Figure 4c shows the temperature dependence of the coherence length $L_\phi$. For $T < 8$ K, $L_\phi$ exceeds the film thickness (20 nm), and we expect that the system is 2D, where the HLN formula is applicable. We expect electron-electron scattering to give a coherence length which has a power-law dependence on temperature $L_\phi \sim T^{-\beta}$ where $\beta = 1/2$ for 2D and $\beta = 3/4$ for 3D systems.[19] The black solid line is a power-law fit to the data. The average, exponent $\beta = 0.86$, is larger than the expected 2D value, and there is some variation in the power-law slope with temperature. However, the data are close to the 2D-3D crossover, and the HLN formula is not strictly applicable in 3D ($T > 8$ K), hence we do not expect perfect agreement. $L_\phi$ reaches 71 nm at 2 K, which is much smaller than as grown films in UHV, where $L_\phi$ are around 300 – 400 nm,[11] consistent with increased disorder.

CONCLUSIONS

In conclusion, we have shown that the Si and $MgF_2$ materials can be used as surface capping layers to protect $Na_3Bi$ thin films from catastrophic degradation in ambient conditions. Our experiments have demonstrated that the electronic properties of $Na_3Bi$ thin films are preserved when it is capped by using Si and $MgF_2$ in UHV and that their resistivity can be kept within metallic region for over 100 hours in ambient. Hence, this capping method makes air-stable $Na_3Bi$ thin film devices possible. Moreover, the analysis of our magnetotransport data of the $MgF_2$-capped $Na_3Bi$ film after in-air exposure shows that the size of the charge puddle region is increased. This suggests that more optimization in the deposition of the capping layer is required. We note that $MgF_2$ is optically transparent and has a high dielectric constant, $\varepsilon$ = 5.289,[20] suggesting future work to achieve both optical measurements and the fabrication of field effect devices utilizing $MgF_2$ as the dielectric media is possible.

FIGURES

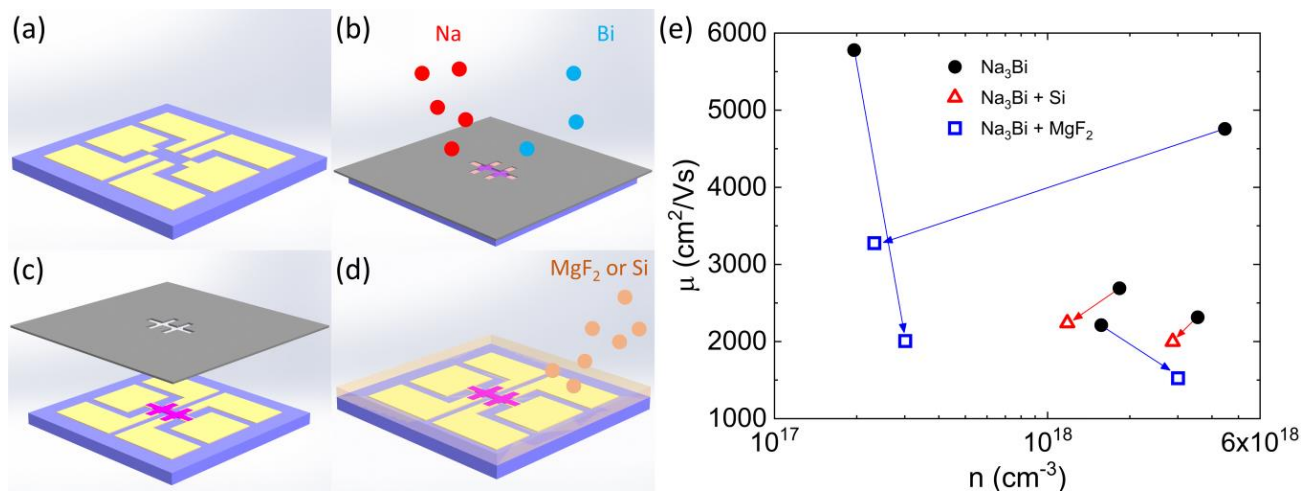

Figure 1. Device fabrication and in-situ Hall effect measurement. Panels (a-d) shows schematically the steps of UHV device fabrication. (a) $Al_2O_3$ substrate with pre-deposited Au contacts. (b) $Na_3Bi$ growth through a shadow mask. (c) Removal of shadow mask after $Na_3Bi$ growth. (d) $MgF_2$ or Si capping layer deposition without shadow mask. (e) Hall carrier density (n) and mobility (µ) of 20 nm thick as-grown (solid circle) $Na_3Bi$ films and Si-capped (hollow triangle) or $MgF_2$-capped (hollow square) $Na_3Bi$ films at 5 K.

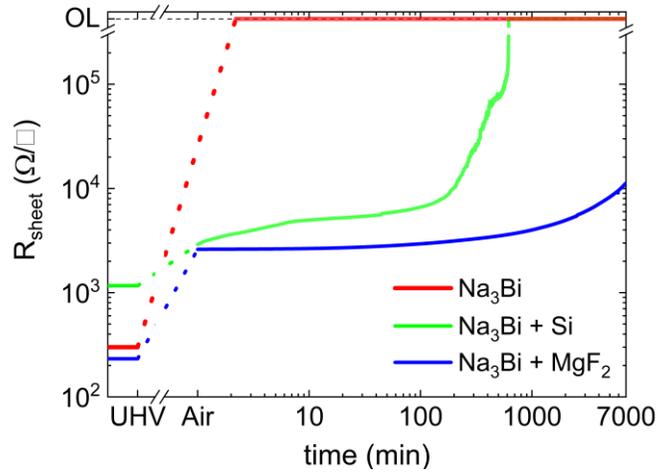

Figure 2. Resistivity of uncapped Na$_3$Bi (solid red line), Si- (solid green line) and MgF$_2$- capped (solid blue line) films as a function of time exposed to ambient atmosphere. Dashed lines are guides to the eye indicating the change in resistance during the transfer from UHV to air during which resistance was not measured. The black dashed line indicates an out-of-range condition for the source meter ($R_{sheet} > 8\times10^6$ Ω/□) in the experimental setup which are further confirmed to be above $10^{10}$ Ω.

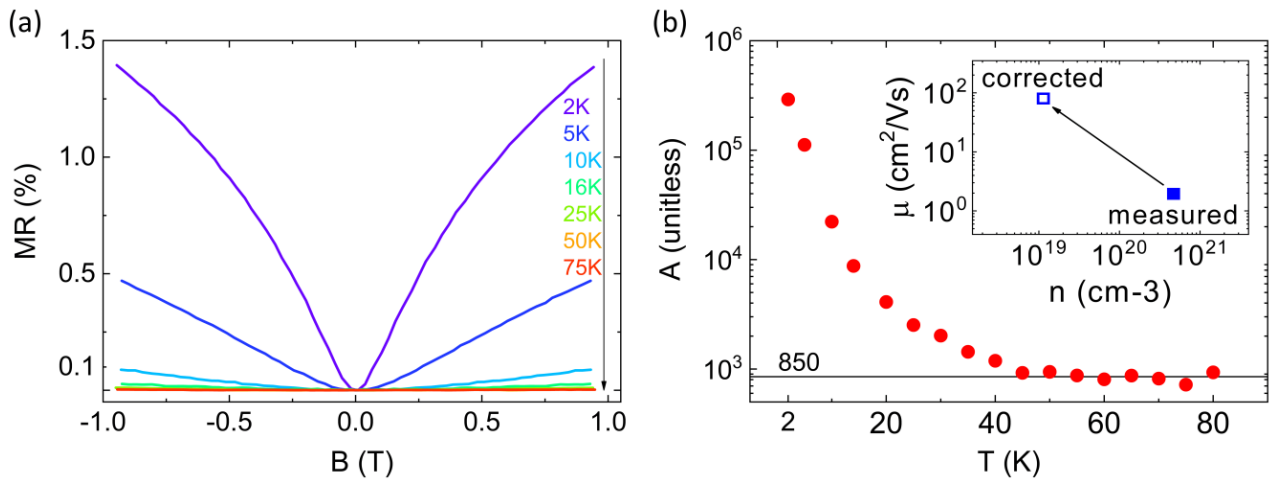

Figure 3. Temperature-dependent magnetotransport characterization of a MgF$_2$ capped Na$_3$Bi film after in-air transfer. (a) Magnetoresistance as a function of magnetic field at temperatures from 2 K to 75 K as indicated in legend. (b) The temperature dependence of the fitting parameter A from Eqn. 1 in main text. The value of A saturates around 850 for T > 50 K. The insert shows the average mobility and carrier density between 50 K and 80 K from the Hall effect measurement (blue solid square), and subsequently corrected for electron-hole puddling effects (blue hollow square; see main text).

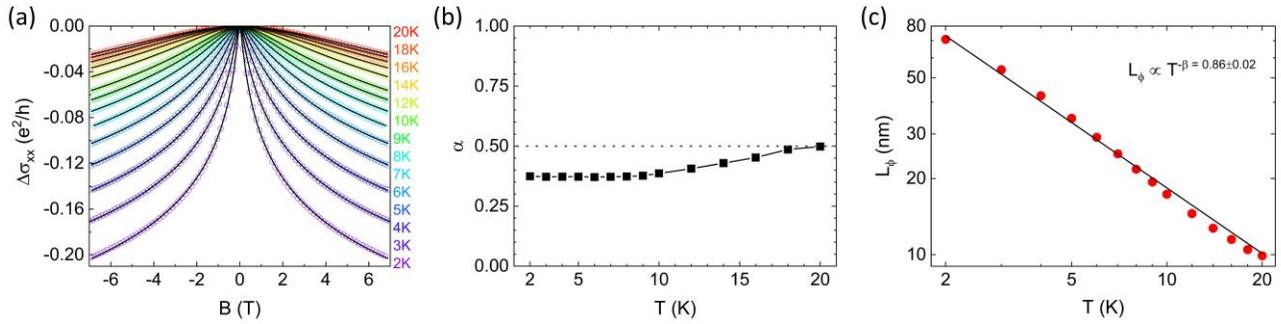

Figure 4. Magnetoconductance of a MgF$_2$ capped Na$_3$Bi film as a function of temperature after in-air transfer. (a) Magnetoconductance as a function of magnetic field at temperatures between 2 K and 20 K (colored circles corresponding to temperature labelled on right). Black solid lines are fits to Eqn. 2 in the main text. (b-c) Temperature dependence of the fit parameters from Eqn. 2 of main text. The pre-factor α is shown in (b) and the phase coherence length L$_\phi$ in (c). The dashed line in (b) where $\alpha = 1/2$ corresponds to the theoretical expectation and the black solid line in (c) indicates a power-law dependence with exponent 0.86±0.02.

## ASSOCIATED CONTENT

**Corresponding Author**

*michael.fuhrer@monash.edu

*mark.edmonds@monash.edu

**Funding Sources**

Australian Research Council Centre for Excellence Future Low Energy Electronics Technologies (CE170100039). ARC Laureate Fellowship (FL120100038). ARC DECRA fellowship (DE160101157). National University of Singapore Young Investigator Award (R-607-000-094-133).

**Notes**

The authors declare no competing financial interest.

ACKNOWLEDGMENT

C.L., J.L.C. and M.S.F., acknowledge the funding support from the Australian Research Council Centre for Excellence Future Low Energy Electronics Technologies (CE170100039). C.L., J.L.C. and M.S.F. are supported by M.S.F.'s ARC Laureate Fellowship (FL120100038). M.T.E. acknowledges the support from ARC DECRA fellowship (DE160101157). S.A. acknowledges the National University of Singapore Young Investigator Award (R-607-000-094-133). We also acknowledge funding from the Monash Centre for Atomically Thin Materials Research Support scheme. This work was performed in part at the Melbourne Centre for Nanofabrication (MCN) in the Victorian Node of the Australian National Fabrication Facility (ANFF).

REFERENCES

(1) Liu, Z.; Zhou, B.; Zhang, Y.; Wang, Z.; Weng, H.; Prabhakaran, D.; Mo, S.-K.; Shen, Z.; Fang, Z.; Dai, X. Discovery of a three-dimensional topological Dirac semimetal, Na3Bi. *Science* **2014,** *343* (6173), 864-867.

(2) Wang, Z.; Sun, Y.; Chen, X.-Q.; Franchini, C.; Xu, G.; Weng, H.; Dai, X.; Fang, Z. Dirac semimetal and topological phase transitions in A 3 Bi (A= Na, K, Rb). *Physical Review B* **2012,** *85* (19), 195320.

(3) Xiong, J.; Kushwaha, S. K.; Liang, T.; Krizan, J. W.; Hirschberger, M.; Wang, W.; Cava, R. J.; Ong, N. P. Evidence for the chiral anomaly in the Dirac semimetal Na3Bi. *Science* **2015,** *350* (6259), 413-416.

(4) Edmonds, M. T.; Collins, J. L.; Hellerstedt, J.; Yudhistira, I.; Gomes, L. C.; Rodrigues, J. N.; Adam, S.; Fuhrer, M. S. Spatial charge inhomogeneity and defect states in topological Dirac semimetal thin films of Na3Bi. *Science advances* **2017,** *3* (12), eaao6661.

(5) Hellerstedt, J.; Yudhistira, I.; Edmonds, M. T.; Liu, C.; Collins, J.; Adam, S.; Fuhrer, M. S. Electrostatic modulation of the electronic properties of Dirac semimetal Na 3 Bi thin films. *Physical Review Materials* **2017,** *1* (5), 054203.

(6) Collins, J. L.; Tadich, A.; Wu, W.; Gomes, L. C.; Rodrigues, J. N.; Liu, C.; Hellerstedt, J.; Ryu, H.; Tang, S.; Mo, S.-K. Electric-field-tuned topological phase transition in ultrathin Na 3 Bi. *Nature* **2018,** *564* (7736), 390-394.

(7) Wray, L. A. Topological transistor. *Nature Physics* **2012,** *8* (10), 705-706.

(8) Pan, H.; Wu, M.; Liu, Y.; Yang, S. A. Electric control of topological phase transitions in Dirac semimetal thin films. *Scientific reports* **2015,** *5*, 14639.

(9) Glinka, Y. D.; Babakiray, S.; Johnson, T. A.; Bristow, A. D.; Holcomb, M. B.; Lederman, D. Ultrafast carrier dynamics in thin-films of the topological insulator Bi2Se3. *Applied Physics Letters* **2013,** *103* (15), 151903.

(10) Bas, D. A.; Muniz, R. A.; Babakiray, S.; Lederman, D.; Sipe, J.; Bristow, A. D. Identification of photocurrents in topological insulators. *Optics express* **2016,** *24* (20), 23583-23595.

(11) Hellerstedt, J.; Edmonds, M. T.; Ramakrishnan, N.; Liu, C.; Weber, B.; Tadich, A.; O'Donnell, K. M.; Adam, S.; Fuhrer, M. S. Electronic properties of high-quality epitaxial topological Dirac semimetal thin films. *Nano letters* **2016,** *16* (5), 3210-3214.

(12) Ping, J.; Yudhistira, I.; Ramakrishnan, N.; Cho, S.; Adam, S.; Fuhrer, M. S. Disorder-induced magnetoresistance in a two-dimensional electron system. *Physical review letters* **2014,** *113* (4), 047206.

(13) Adam, S.; Hwang, E.; Galitski, V.; Sarma, S. D. A self-consistent theory for graphene transport. *Proceedings of the National Academy of Sciences* **2007,** *104* (47), 18392-18397.

(14) Martin, J.; Akerman, N.; Ulbricht, G.; Lohmann, T.; Smet, J. v.; Von Klitzing, K.; Yacoby, A. Observation of electron–hole puddles in graphene using a scanning single-electron transistor. *Nature physics* **2008,** *4* (2), 144-148.

(15) Zhang, Y.; Brar, V. W.; Girit, C.; Zettl, A.; Crommie, M. F. Origin of spatial charge inhomogeneity in graphene. *Nature Physics* **2009,** *5* (10), 722-726.

(16) Xue, J.; Sanchez-Yamagishi, J.; Bulmash, D.; Jacquod, P.; Deshpande, A.; Watanabe, K.; Taniguchi, T.; Jarillo-Herrero, P.; LeRoy, B. J. Scanning tunnelling microscopy and spectroscopy of ultra-flat graphene on hexagonal boron nitride. *Nature materials* **2011,** *10* (4), 282-285.

(17) Beidenkopf, H.; Roushan, P.; Seo, J.; Gorman, L.; Drozdov, I.; San Hor, Y.; Cava, R. J.; Yazdani, A. Spatial fluctuations of helical Dirac fermions on the surface of topological insulators. *Nature Physics* **2011,** *7* (12), 939.

(18) Hikami, S.; Larkin, A. I.; Nagaoka, Y. Spin-orbit interaction and magnetoresistance in the two dimensional random system. *Progress of Theoretical Physics* **1980,** *63* (2), 707-710.

(19) Altshuler, B. L.; Aronov, A.; Khmelnitsky, D. Effects of electron-electron collisions with small energy transfers on quantum localisation. *Journal of Physics C: Solid State Physics* **1982,** *15* (36), 7367.

(20) Fontanella, J.; Andeen, C.; Schuele, D. Low-frequency dielectric constants of α-quartz, sapphire, MgF2, and MgO. *Journal of Applied Physics* **1974,** *45* (7), 2852-2854.